\documentclass[12pt]{article}
\usepackage{amsmath,amssymb,mathrsfs,url,graphicx,hyperref}
\usepackage{color}

\title{Absorbing Boundary Condition as Limiting Case of Imaginary Potentials}

\author{
Roderich Tumulka\footnote{Fachbereich Mathematik, Eberhard-Karls-Universit\"at, Auf der Morgenstelle 10, 72076 T\"ubingen, Germany. E-mail: roderich.tumulka@uni-tuebingen.de}
}

\date{November 28, 2019}

\newcommand{\be}{\begin{equation}}
\newcommand{\ee}{\end{equation}}
\newcommand{\scp}[2]{\langle #1|#2\rangle}
\renewcommand{\Re}{\mathrm{Re}}
\renewcommand{\Im}{\mathrm{Im}}

\newcommand{\RRR}{\mathbb{R}}
\newcommand{\CCC}{\mathbb{C}}

\newcommand{\vX}{{\boldsymbol{X}}}

\newcommand{\prob}{\mathrm{Prob}}

\begin{document}
\maketitle
\begin{abstract}
Imaginary potentials such as $V(x)=-iv 1_\Omega(x)$ (with $v>0$ a constant, $\Omega$ a subset of 3-space, and $1_\Omega$ its characteristic function) have been used in quantum mechanics as models of a detector. They represent the effect of a ``soft'' detector that takes a while to notice a particle in the detector volume $\Omega$. In order to model a ``hard'' detector (i.e., one that registers a particle as soon as it enters $\Omega$), one may think of taking the limit $v\to\infty$ of increasing detector strength $v$. However, as pointed out by Allcock, in this limit the particle never enters $\Omega$; its wave function gets reflected at the boundary $\partial \Omega$ of $\Omega$ in the same way as by a Dirichlet boundary condition on $\partial \Omega$. This phenomenon, a cousin of the ``quantum Zeno effect,'' might suggest that a hard detector is mathematically impossible. Nevertheless, a mathematical description of a hard detector has recently been put forward in the form of the ``absorbing boundary rule'' involving an absorbing boundary condition on the detecting surface $\partial \Omega$. We show here that in a suitable (non-obvious) limit, the imaginary potential $V$ yields a non-trivial distribution of detection time and place in agreement with the absorbing boundary rule. That is, a hard detector can be obtained as a limit, but it is a different limit than Allcock considered.

\medskip

\end{abstract}

\section{Introduction}

Imaginary potentials have the effect that the time evolution defined by the Schr\"odinger equation is no longer unitary; rather, they lead to gain or loss of $|\psi|^2$ weight, depending on whether the potential is positive or negative imaginary. Such a loss is desirable to model absorption or detection of particles \cite{Bet40,MM65,Lev69,All69b,Hod71}. Here, we are interested in detection and consider two kinds of detectors: a ``hard'' detector that registers a particle as soon as it enters the detectors volume, and a ``soft'' detector that takes some time to register the particle. Imaginary potentials are suitable as models of a soft detector, as discussed in particular by Allcock \cite{All69b}. For example, for a single, non-relativistic quantum particle of mass $m>0$ in 1 dimension with a soft detector in the region $[0,\infty)$, we consider the Schr\"odinger equation
\be\label{SchriV}
i\hbar \frac{\partial \psi}{\partial t} = -\tfrac{\hbar^2}{2m} \frac{\partial^2 \psi}{\partial x^2} - iv\, \Theta(x) \, \psi(x)\,,
\ee
where $v>0$ is a constant (the detection rate) and $\Theta$ is the Heaviside function [i.e., $\Theta(x)=1$ for $x\geq 0$ and 0 otherwise]. 

It is less obvious how to model a hard detector. In particular, one might expect that such a model could be obtained as a limit of a soft detector, letting the parameter $v$, representing the strength of the detector, tend to $\infty$. However, Allcock \cite{All69b} found that in this limit, for an initial wave function concentrated in the negative half axis, with probability 1 the detector never clicks, and $\psi_t(x)=0$ at all $x\geq 0$ and $t\geq 0$---a situation reminiscent of the quantum Zeno paradox \cite{All69b,Fri72,Zeno,Dhar13,Dhar14}. Allcock (prematurely) concluded that a hard detector was mathematically impossible. 

A successful proposal for modeling a hard detector is provided by the ``absorbing boundary rule'' \cite{Wer87,detect-rule}. According to it, the wave function $\psi_t:(-\infty,0]\to \CCC$ evolves according to the free Schr\"odinger equation
\be\label{Schr}
i\hbar \frac{\partial \psi}{\partial t} = -\tfrac{\hbar^2}{2m} \frac{\partial^2 \psi}{\partial x^2}\,,
\ee
supplemented by the ``absorbing'' boundary condition
\be\label{abc}
\frac{\partial \psi_t}{\partial x}(0) = i\kappa \psi_t(0)\,,
\ee
where $\kappa>0$ is a constant (the wave number of sensitivity of the detector). For other proposed rules about the detection time distribution, see \cite{AB61,Kij74,GRT96,AOPRU98,Bau00,ML00,MSE08,MRC09}.

In this note, we describe a limiting procedure, different from the limit $v\to\infty$ that we will henceforth call ``Allcock's limit,'' in which the soft detector model \eqref{SchriV} approaches the hard detector model given by the absorbing boundary rule \eqref{Schr}, \eqref{abc}. Our derivations are not mathematically rigorous. The convergence occurs for wave functions as well as for the distribution of the detection time, specified in \eqref{PTiV} in Section~\ref{sec:setup}. 

We proceed as follows. Both the Hamiltonian $H_{iv}$ of \eqref{SchriV} with imaginary potential and the Hamiltonian $H^{i\kappa}$ of \eqref{Schr}, \eqref{abc} with absorbing boundary condition are non-selfadjoint, and the time evolution operators $W_t=\exp(-iHt/\hbar)$ they define are not unitary but are contractions (i.e., $\|W_t\psi\|\leq \|\psi\|$). We compute the eigenfunctions and eigenvalues of the Hamiltonians. Since the Hamiltonians are not selfadjoint, either their eigenvalues are complex and not all real, or their eigenfunctions are not all mutually orthogonal; both situations occur in various cases. We want to show that the eigenfunctions and eigenvalues of $H_{iv}$ approach those of $H^{i\kappa}$ in a suitable limit. This limit involves one more modification of $H_{iv}$: we make the detector volume a finite interval $[0,L]$ and impose Neumann boundary conditions at $L>0$,
\be\label{NeuL}
\frac{\partial \psi_t}{\partial x}(L)=0\,.
\ee
The (non-selfadjoint) Hamiltonian in $L^2\bigl((-\infty,L]\bigr)$ defined by \eqref{SchriV} and \eqref{NeuL} will be denoted by $H_{iv,L}$. We claim that
\be\label{lim}
H_{iv,L}\to H^{i\kappa}
\ee
in the
\be\label{hardlim}
\text{``hard'' limit}\quad L\to 0, \quad v\to \infty, \quad vL \to \frac{\hbar^2\kappa}{2m}>0 
\ee
while keeping $m$ (and $\hbar$) constant. Moreover, the distribution of the detection time converges to that of the absorbing boundary rule in the hard limit.

Analogous Hamiltonians on a lattice, along with similar questions of convergence and of avoiding the quantum Zeno effect, are considered in \cite{detect-lattice}.

In Section~\ref{sec:setup}, we give more detail about the models of hard and soft detectors. In Section~\ref{sec:eig}, we compute their eigenvalues and eigenfunctions. In Section~\ref{sec:limit}, we derive the limiting statement. In Section~\ref{sec:rem}, we conclude with some remarks.

\section{Setup of Equations}
\label{sec:setup}

\subsection{Imaginary Potential} 

The Schr\"odinger equation \eqref{SchriV} with imaginary potential leads to the continuity equation
\be\label{contiV}
\frac{\partial |\psi|^2}{\partial t} = -\frac{\partial j}{\partial x} -\tfrac{2v}{\hbar} \Theta(x) \, |\psi(x)|^2
\ee
with probability current
\be\label{jdef}
j= \tfrac{\hbar}{m} \Im \Bigl[\psi^* \frac{\partial \psi}{\partial x}\Bigr].
\ee
To visualize the physical meaning of \eqref{contiV}, we may think of the Bohmian trajectory associated with it: It is the solutions $t\mapsto X(t)$ of the equation of motion $dX/dt=j(X)/|\psi(X)|^2$ that has random initial condition $X(0)$ with $|\psi_0|^2$ distribution and ends at a random time with rate $(2v/\hbar) \Theta(X(t))$; that is, whenever the particle is in the detector volume, it has probability $(2v/\hbar)dt$ to disappear in the next $dt$ seconds. We can think of this disappearance as an absorption due to detection. As a consequence, assuming $\|\psi_0\|=1$, the probability distribution of the time $T$ and place $X$ of detection (or, equivalently, of the end of the trajectory) is
\be\label{PTiV}
\prob\Bigl(t_1\leq T \leq t_2, X\in B\Bigr) = \int_{t_1}^{t_2}dt \int_B dx \, \tfrac{2v}{\hbar}|\psi_t(x)|^2\,,
\ee
along with the probability
\be\label{PinftyiV}
\prob(T=\infty) = \lim_{t\to\infty} \int_{\RRR} dx \, |\psi_t(x)|^2
\ee
that the particle nevers gets detected (as could happen, for example, if the particle wanders off to $-\infty$ without ever entering the detector volume). The quantity $|\psi_t(x)|^2\, dx$ represents the probability that the particle is located in $[x,x+dx]$ at time $t$ (and, in particular, has not been absorbed yet). It follows that
\be
\|\psi_t\|^2 = \prob(T>t)
\ee 
is the ``survival probability,'' and since $\prob(t<T<t+dt) = \prob(T>t) - \prob(T>t+dt)$, that the probability density of $T$ is
\be\label{rhoTpsitiV}
\rho_T(t) = -\frac{d}{dt} \|\psi_t\|^2\,.
\ee
Should the experiment be terminated at a time $t$ before the detector clicks, then the collapsed wave function is $\psi_t/\|\psi_t\|$.

\subsection{Absorbing Boundary Rule}

It is known \cite{Wer87,detect-thm}, that the system \eqref{Schr}--\eqref{abc}  possesses a unique solution $\psi_t(x)$ for every initial datum $\psi_0\in L^2\bigl((-\infty,0]\bigr)$; we will assume $\|\psi_0\|=1$. The probability distribution of the random time $T$ at which the detector clicks has density $\rho_T(t)$ given by the probability current $j_t$ at $x=0$, that is,
\be\label{rhoTABR}
\rho_T(t) = \tfrac{\hbar}{m} \Im\Bigl[ \psi_t^*(0) \frac{\partial \psi_t}{\partial x}(0)  \Bigr] \,. 
\ee
By virtue of the boundary condition \eqref{abc}, this quantity can equivalently be expressed as
\be
\rho_T(t) = \tfrac{\hbar\kappa}{m} |\psi_t(0)|^2\,,
\ee
which is clearly non-negative, as a probability density must be; we see in particular that the current $j_t(0)$ is always pointing outward. 
Again, $|\psi_t(x)|^2dx$ is the probability for the presence of the particle in $[x,x+dx]$ at time $t$, and the distribution of $T$ can equivalently be rewritten as
\be\label{rhoTpsitABR}
\rho_T(t) = -\frac{d}{dt} \|\psi_t\|^2\,.
\ee
Moreover, again, should the experiment be terminated at a time $t$ before the detector clicks, then the collapsed wave function is $\psi_t/\|\psi_t\|$.
 
Extensions of the absorbing boundary rule to moving detectors and to several particles are described in \cite{detect-several}, and to the Dirac equation in \cite{detect-dirac}. 
Note also that the theory still works in the same way if we replace the boundary condition \eqref{abc} by
\be\label{bc2}
\frac{\partial \psi_t}{\partial x}(0) = (\nu+i\kappa) \psi_t(0)
\ee
with arbitrary $\nu\in \RRR$ (and still $\kappa>0$). The Hamiltonian $H^{\nu+i\kappa}$ is then defined as $-(\hbar^2/2m)\partial^2/\partial x^2$ with boundary condition \eqref{bc2}.

\section{Eigenvalues and Eigenfunctions}
\label{sec:eig}

\subsection{Hamiltonian with Imaginary Potential}
\label{sec:eigiV}

We aim at finding the eigenvalues and (non-normalizable) eigenfunctions of $H_{iv,L}$ defined by \eqref{SchriV} and \eqref{NeuL}. 

Focus first on $x<0$ (``region I''); being an eigenfunction of $-(\hbar^2/2m)\partial^2/\partial x^2$ means to solve an ODE in $x$ whose general solution has the form
\be\label{fI1}
f_I(x) = d_k \, e^{ikx} + c_k \, e^{-ikx} \,,
\ee
possibly with complex $k$. It seems plausible that, although $H_{iv,L}$ will not be self-adjoint, only real $k$ are relevant to the eigenfunctions, as $\exp(ikx)$ with $k>0$ then represents an incoming wave from the left. It follows that
\be\label{Ek}
E=\frac{\hbar^2k^2}{2m}\,,
\ee
and we can choose without loss of generality that $k>0$, so that $k=\sqrt{2mE}/\hbar$; we can and will also choose $d_k=1$, so
\be\label{fI2}
f_I(x) = e^{ikx} + c_k \, e^{-ikx} \,.
\ee

Let us turn to $0<x<L$ (``region II''). Any eigenfunction must then have the form
\be\label{fII}
f_{II}(x) = a_k \, e^{i\lambda x} + b_k \, e^{-i\lambda x}
\ee
with complex $\lambda$ satisfying
\be\label{lambdadef}
\lambda^2 = k^2 + i \frac{2mv}{\hbar^2}
\ee
and, say, $\Re \, \lambda>0$ (to define which of the two square roots is called $\lambda$ and which $-\lambda$). The contribution that shrinks exponentially may seem plausible in view of the absorption taking place in region II; the contribution that grows exponentially may be thought of as reflected at $L$.

It also seems plausible that the eigenfunctions should satisfy matching conditions
\be\label{match}
f_I(0)=f_{II}(0)\quad \text{and} \quad \frac{\partial f_I}{\partial x}(0) = \frac{\partial f_{II}}{\partial x}(0)\,,
\ee
which imply that
\be\label{cab}
1+c_k = a_k + b_k~~~\text{and}~~~k(1-c_k) = \lambda (a_k-b_k)\,.
\ee
The Neumann boundary condition \eqref{NeuL} at $L$ implies that 
\be\label{bLa}
b_k=e^{i2\lambda L}a_k\,.
\ee
From these three relations \eqref{cab}, \eqref{bLa} together, we obtain that
\be\label{cklambda}
c_k = \frac{(k-\lambda)+(k+\lambda) e^{i2\lambda L}}{(k+\lambda) + (k-\lambda)e^{i2\lambda L}}\,.
\ee
We also note for later use the explicit expression for $a_k$,
\be\label{aklambda}
a_k = \frac{2k}{(k+\lambda)+(k-\lambda)e^{i2\lambda L}}\,.
\ee

\subsection{Hamiltonian with Absorbing Boundary Condition}
\label{sec:eigabc}

We now aim at finding the eigenvalues and eigenfunctions of $H^{\nu+i\kappa}$, defined by the Schr\"odinger equation \eqref{Schr} and the absorbing boundary condition in the more general version \eqref{bc2}. 

Again, eigenfunctions must be of the form \eqref{fI1}, and again, $k$ should be real, without loss of generality positive, and $d_k$ can be taken to be 1, so the eigenfunction is again of the form \eqref{fI2} with real eigenvalue given by \eqref{Ek}. This time, the coefficient
$c_k\in\CCC$ must be chosen so that \eqref{bc2} is satisfied, i.e., $ik(1-c_k) = (\nu+i\kappa)(1+c_k)$ or
\be\label{ck}
c_k = \frac{k-\kappa+i\nu}{k+\kappa-i\nu}\,.
\ee
In this case, there is no region II.

We remark that the eigenfunctions are not mutually orthogonal. A formal calculation of their inner product yields that, for $k\neq k'$ and $f_k$ the eigenfunction associated with $k>0$,
\be\label{scpff}
\scp{f_{k'}}{f_{k}} 
= -i \frac{1-c_{k'}^*c_k}{k-k'} - i \frac{c_{k'}^*-c_k }{k+k'} \,,
\ee
which in general will not vanish.
Indeed, starting from the equation $f''_k=-k^2 f_k$ and the conjugate equation for $k'$, multiply by $f^*_{k'}$ respectively $f_k$, then subtract the two equations, then integrate over $x$ from $-\infty$ to 0. Integrating by parts on the left hand side, the integral on the left vanishes. Neglecting boundary terms at $x=-\infty$, we obtain that 
\be
f^*_{k'}(0) f'_k(0) - f^{\prime *}_{k'}(0) f_k(0) 
= (-k^2 + k^{\prime 2}) \int\limits_{-\infty}^0 dx\,  f^*_{k'}(x)\,  f_k(x)\,,
\ee
and the explicit form \eqref{fI2} then yields \eqref{scpff}.

\subsection{Reflection Coefficient}
\label{sec:refl}

The eigenfunctions $f(x)$ contain a right-moving wave $e^{ikx}$ ($k>0$) coming from $-\infty$ and a reflected wave $c_k e^{-ikx}$ coming from the right boundary (at 0) and moving to the left. The absolute square of $c_k$ provides the reflection coefficient \cite{LL}
\be
R_k=|c_k|^2
\ee
or idealized probability of reflection at this value of $k$; the absorption coefficient is $A_k=1-R_k$. As discussed in \cite{detect-rule}, an \emph{absorbing boundary} means that the particle gets absorbed there, but not necessarily (or not completely) the wave. Perfect absorption, $A_k=1$, is reached when $R_k=0$ or $c_k=0$, and by \eqref{ck} this occurs for $H^{\nu+i\kappa}$ whenever
\be\label{Rk0}
k-\kappa+i\nu =0\,.
\ee
Since $k$ is real, this situation can only occur when $\nu=0$, and that is why an \emph{ideal} detector was assumed to have $\nu=0$ in \cite{detect-rule}.

\section{Limiting Cases}
\label{sec:limit}

\subsection{Allcock's Limit}

The simplest situation in which we can consider Allcock's limit has no right boundary, $L=\infty$. The eigenvalues and eigenfunctions for this situation can actually be obtained from the formulas of Section~\ref{sec:eigiV} in the limit $L\to\infty$. Fix $v>0$ and $k>0$; since $\lambda^2$ has phase between 0 and $\pi/2$ and thus $\lambda$ between 0 and $\pi/4$, we know that $\lambda$ has positive imaginary part, with the consequence that as $L\to\infty$, $e^{i\lambda L}\to 0$ and $e^{i2\lambda L}\to 0$. Thus, $b_k\to 0$, so the exponentially growing contribution to $f_{II}$ disappears and, in the limit $L\to\infty$,
\be
c_k = \frac{k-\lambda}{k+\lambda}\,.
\ee

Now take Allcock's limit $v\to\infty$. Since $|\lambda^2|\to\infty$, also $|\lambda|\to\infty$, so
\be
c_k \to -1\,.
\ee
In particular, the reflection coefficient is 1 and the absorption coefficient 0. As Allcock found, the probability that the particle ever gets detected (and thus absorbed) is 0.

In fact, in the limit $v\to\infty$ also $a_k\to 0$, so $f_{II}(x)\to 0$ for every $x>0$, so the probability of the particle ever entering the detector volume is 0. Since $c_k\to -1$, 
\be
f_I(x) \to e^{ikx} - e^{-ikx}\,,
\ee
which are the eigenfunctions of the Schr\"odinger equation on $(-\infty,0]$ with a Dirichlet boundary condition
\be\label{Dir}
\psi_t(0)=0.
\ee

\subsection{Hard Limit}

We now consider the hard limit \eqref{hardlim} and show that $H_{iv,L}\to H^{i\kappa}$ with $\nu=0$ in the sense that the eigenfunctions and eigenvalues converge.

Our first claim is that in this limit, for any $k>0$,
\be\label{limit1}
\frac{1-e^{i2\lambda L}}{1+e^{i2\lambda L}}\lambda \to \kappa>0\,.
\ee
Indeed, $\lambda^2\to i\infty$, $\lambda\to (1+i)\infty$, $\lambda^2 L \to i\kappa$, $\lambda L\to 0$, $e^{i2\lambda L}\to 1$, and $(1-e^{i2\lambda L})\lambda \approx (-i2\lambda L) \lambda=-i2 \lambda^2 L \to 2\kappa$, which implies \eqref{limit1}. 

Now \eqref{cklambda} can be rewritten as
\be\label{ck3}
c_k 
= \frac{(1+e^{i2\lambda L})k-(1-e^{i2\lambda L})\lambda}{(1+e^{i2\lambda L})k+(1-e^{i2\lambda L})\lambda}
= \frac{k-\frac{1-e^{i2\lambda L}}{1+e^{i2\lambda L}}\lambda}{k+\frac{1-e^{i2\lambda L}}{1+e^{i2\lambda L}}\lambda}\,,
\ee
and from \eqref{limit1} it follows that
\be\label{ck4}
c_k \to \frac{k-\kappa}{k+\kappa}\,,
\ee
which agrees with \eqref{ck}, the $c_k$ of the absorbing boundary condition \eqref{bc2} with $\nu=0$. Thus, $f_I$ converges to the $f_I$ of $H^{i\kappa}$. 

In order to show that for any given $k>0$, the eigenfunction of $H_{iv,L}$ converges to that of $H^{i\kappa}$, we need to verify that $f_{II}$ disappears in the hard limit. While the interval $[0,L]$ shrinks to a point, it is not as obvious that
\be
\|f_{II}\|^2=\int_0^L dx \, |f_{II}(x)|^2 \to 0\,.
\ee
To see that this is indeed the case, note that, by \eqref{aklambda} and the relations mentioned between \eqref{limit1} and \eqref{ck3},
\be
a_k = \frac{2k}{(1+e^{i2\lambda L})k+(1-e^{i2\lambda L})\lambda} \to\frac{k}{k+\kappa}
\ee
as well as, by \eqref{bLa}, $b_k \to k/(k+\kappa)$. Hence,
\begin{align}
\|b_k e^{-i\lambda x}\|^2 
&= |b_k|^2 \int_0^L dx \, |e^{-i\lambda x}|^2\\
&= |b_k|^2 \int_0^L dx \, e^{2\Im \lambda x}\\
&= |b_k|^2 \int_0^L dx \, e^{4mvx/\hbar^2}\\
&= |b_k|^2 \, \frac{e^{4mvL/\hbar^2}-1}{4mv/\hbar^2}\\[2mm]
&\to 0
\end{align}
since $b_k$ stays bounded, $vL\to \hbar^2\kappa/2m$ stays bounded, and $v\to\infty$. In a similar way one can see that also $\|a_k e^{i\lambda x}\|\to 0$, so that $\|f_{II}\|\to 0$, as claimed.

We thus have that the eigenfunctions converge, while the eigenvalue is the same, viz., \eqref{Ek}. That is, $H_{iv,L}\to H^{i\kappa}$ in the hard limit. It is thus also plausible that $\exp(-iH_{iv,L}t/\hbar)\to \exp(-iH^{i\kappa}t/\hbar)$, and that $\psi_t$ converges accordingly for every fixed $t$.

Our further claim that the distribution density $\rho_T$ of the detection time $T$ converges to that of the absorbing boundary rule then follows from the fact that in both settings (the imaginary potential and the absorbing boundary), $\rho_T(t)=-d\|\psi_t\|^2/dt$, see \eqref{rhoTpsitiV} and \eqref{rhoTpsitABR}.

\section{Remarks}
\label{sec:rem}

\begin{enumerate}
\item {\it Higher dimension.} Our analysis of the hard limit carries over directly to the case in which the detector volume is $\{(x_1,\ldots,x_d):0<x_1<L\}$, the particle is restricted to $x_1<L$, and a Neumann boundary condition is imposed at $x_1=L$. The question then arises whether also the probability distribution of the detection place $\vX$ (and the joint distribution of $T$ and $\vX$) obtained from the imaginary potential model converges to that obtained from the absorbing boundary rule \cite{detect-rule}. That this should be so is visible in the Bohmian picture: If $\psi_t$ for the imaginary potential converges to $\psi_t$ for the absorbing boundary, then also the Bohmian trajectories should converge. Since in the imaginary potential case, the particle can only be absorbed if $X(t)>0$, in the limit the particle can only be absorbed when reaching $x=0$; but then it \emph{must} be absorbed since there is no Bohmian trajectory leading from the boundary to the left. So in the limit all trajectories must end exactly when they reach the boundary, so the distribution of the detection events must agree with the distribution of the arrival events \cite{Daumer}, which coincides with the distribution \eqref{rhoTABR} of the detection events according to the absorbing boundary rule.

\item {\it General shapes in higher dimension.} It seems plausible that the hard limit still yields the absorbing boundary rule for more general shapes of the detecting surface, as the limit $L\to 0$ focuses on small length scales, on which a curved surface looks flat; also surfaces with edges (such as that of a cube) seem unproblematical since the probability current into the edges will be negligible. It also seems very plausible that the joint probability distribution of the detection time $T$ and the detection location $\vX$ approaches that of the absorbing boundary rule.

\item {\it Robin condition.} The hard limit still agrees with the absorbing boundary rule if we replace the Neumann condition \eqref{NeuL} by a Robin condition
\be\label{Robin}
\frac{\partial \psi}{\partial x}(L)=\alpha\,\psi(L)
\ee
with arbitrary constant $\alpha\in\RRR$, but not if we replace it by a Dirichlet condition $\psi(L)=0$. That is because for a Robin condition, the factor $\exp(i2\lambda L)$ gets replaced by $\frac{i\lambda-\alpha}{i\lambda +\alpha}\exp(i2\lambda L)$, and any limit involving $v\to \infty$ entails that $\lambda \to (1+i)\infty$, so $\frac{i\lambda-\alpha}{i\lambda +\alpha}\to 1$, and the limiting behavior of \eqref{cklambda} is the same as in the Neumann case $\alpha=0$. In the Dirichlet case, however, $\exp(i2\lambda L)$ gets replaced by $-\exp(i2\lambda L)$, and \eqref{limit1} cannot possibly hold with the opposite signs because the growth of $\lambda$ requires, in view of the bounded denominator in \eqref{limit1}, that the numerator $1+\exp(i2\lambda L)$ tends to 0, but that cannot occur because $\lambda L$ must approach a positive multiple of $1+i$. However, \eqref{limit1} is necessary, in view of \eqref{ck3}, for \eqref{ck4} to hold.

\item {\it Finite interval.}
Consider now a finite interval, which it will be convenient to take to be $[-\ell,0]$, with the absorbing boundary condition \eqref{bc2} at 0 and, for example, a Dirichlet boundary condition at $-\ell$, $\psi(-\ell)=0$. Then the eigenfunctions are still of the form \eqref{fI2}, but they need to satisfy in addition $e^{-ik\ell}+ c_k e^{ik\ell}=0$ or
\be\label{kcond}
\frac{k-\kappa+i\nu}{k+\kappa-i\nu} =-e^{-i2k\ell}\,,
\ee
which restricts the possible $k$ values to a discrete set and forces them to become complex, resulting in complex eigenvalues \eqref{Ek} with negative imaginary parts. The same consequence, discrete complex eigenvalues, would occur for the imaginary potential as in \eqref{SchriV} on the interval $[-\ell,L]$ with boundary conditions at $-\ell$ and $L$ such as Dirichlet at $-\ell$ and Neumann at $L$. The eigenfunctions $f$ are now square-integrable, and the complex eigenvalues $\omega=E-i\mu$ (with $E\in\RRR,\mu>0$) ensure that
\be
\|f_t\|^2=\bigl\|\exp(-iHt/\hbar)f\bigr\|^2=\bigl\|\exp(-i\omega t/\hbar)f\bigr\|^2=\exp(-2\mu t/\hbar) \, \|f\|^2
\ee
shrinks with time. I expect that also in this situation the Hamiltonian with imaginary potential converges in the hard limit \eqref{hardlim} (keeping $\ell$ constant) to the one with absorbing boundary condition with $\nu=0$; a careful study of this question would be of interest.
\end{enumerate}

\bigskip

\noindent \textit{Acknowledgments.} I thank Stephen Shipman for helpful discussions.


\begin{thebibliography}{28.}

\bibitem{AB61} Y. Aharonov and D. Bohm:
	Time in the Quantum Theory and the Uncertainty Relation for Time and Energy.
	{\it Physical Review} {\bf 122}: 1649 (1961)

\bibitem{AOPRU98} Y. Aharonov, J. Oppenheim, S. Popescu, B. Reznik, and W.G. Unruh:
	Measurement of time of arrival in quantum mechanics.
	{\it Physical Review A} {\bf 57}: 4130 (1998)
	\url{http://arxiv.org/abs/quant-ph/9709031}


\bibitem{All69b} G.R. Allcock:
	The time of arrival in quantum mechanics II. The individual measurement.
	{\it Annals of Physics} {\bf 53}: 286--310 (1969)

\bibitem{Bau00} A.D. Baute:
	Time-of-arrival distributions from position-momentum and energy-time joint measurements.
	{\it Physical Review A} {\bf 61}: 521111 (2000)

\bibitem{Bet40} H.A. Bethe:
	A Continuum Theory of the Compound Nucleus.
	{\it Physical Review} {\bf 57}: 1125--1144 (1940)

\bibitem{Daumer} M. Daumer, D. D\"urr, S. Goldstein, and N. Zangh\`\i:
	On the quantum probability flux through surfaces.
	\textit{Journal of Statistical Physics} \textbf{88}: 967--977 (1997).
	\url{http://arxiv.org/abs/quant-ph/9512016}

\bibitem{Dhar13} S. Dhar, S. Dasgupta, and A. Dhar:
	Quantum time of arrival distribution in a simple lattice model.
	\textit{Journal of Physics A: Mathematical and Theoretical} \textbf{48}: 115304 (2015)
	\url{http://arxiv.org/abs/1312.5923}

\bibitem{Dhar14} S. Dhar, S. Dasgupta, A. Dhar, and D. Sen:
	Detection and Survival of a Quantum Particle on a Lattice.
	{\it Physical Review A} {\bf 91}: 062115 (2015)
	\url{http://arxiv.org/abs/1410.8701}

\bibitem{detect-lattice} A. Dhar, S. Teufel, and R. Tumulka:
	Detection Time Distribution in Quantum Mechanics on a Lattice.
	In preparation (2020)

\bibitem{Fri72} C.N. Friedman:
	Semigroup product formulas, compressions, and continual observations 
	in quantum mechanics.
	{\it Indiana Mathematical Journal} {\bf 21}: 1001--1011 (1972)

\bibitem{GRT96} N. Grot, C. Rovelli, and R.S. Tate:
	Time-of-arrival in quantum mechanics.
	{\it Physical Review A} {\bf 54}: 4676 (1996)
	\url{http://arxiv.org/abs/quant-ph/9603021}

\bibitem{HSM03} G.C. Hegerfeldt, D. Seidel, and J.G. Muga:
	Quantum arrival times and operator normalization.
	{\it Physical Review A} {\bf 68}: 022111 (2003)
	\url{http://arxiv.org/abs/quant-ph/0308087}

\bibitem{Hod71} P.E. Hodgson:
	{\it Nuclear Reactions and Nuclear Structure}.
	Oxford: Clarendon (1971)

\bibitem{Kij74} J. Kijowski:
	On the time operator in quantum mechanics and the Heisenberg 
	uncertainty relation for energy and time.
	{\it Reports on Mathematical Physics} {\bf 6}: 361 (1974)

\bibitem{LL} L.D. Landau and E.M. Lifshitz:
	{\it Quantum Mechanics, Non-Relativistic Theory. Vol.~3 of Course of Theoretical Physics.} 
	3rd edition, translated from the Russian by J.B. Sykes and J.S. Bell. 
	Oxford: Pergamon (1977)


\bibitem{Lev69} R.D. Levine:
	{\it Quantum Mechanics of Molecular Rate Processes.}
	Oxford University Press (1969)


\bibitem{Zeno} B. Misra and E.C.G. Sudarshan: 
	The Zeno's paradox in quantum theory.
	\textit{Journal of Mathematical Physics} \textbf{18}: 756--763 (1977)

\bibitem{MM65} N.F. Mott and H.S. Massey:
	{\it The Theory of Atomic Collisions}, 3rd edition.
	Oxford University Press (1965)

\bibitem{ML00} J.G. Muga and R. Leavens:
	Arrival Time in Quantum Mechanics.
	\textit{Physics Reports} \textbf{338}: 353 (2000)

\bibitem{MSE08} J.G. Muga, R. Sala Mayato, \'I.L. Egusquiza (editors):
	{\it Time in Quantum Mechanics, Vol.~1, Second Edition.}
	Lecture Notes in Physics {\bf 734}.
	New York: Springer (2008)

\bibitem{MRC09} J.G. Muga, A. Ruschhaupt, A. del Campo (editors):
	{\it Time in Quantum Mechanics, Vol.~2.}
	Lecture Notes in Physics {\bf 789}.
	New York: Springer (2009)

\bibitem{detect-thm} S. Teufel and R. Tumulka:
	Existence of Schr\"odinger Evolution with Absorbing Boundary Conditions.
	In preparation (2019)

\bibitem{detect-rule} R. Tumulka:
	Distribution of the Time at Which an Ideal Detector Clicks.
	Preprint (2016)
	\url{http://arxiv.org/abs/1601.03715}

\bibitem{detect-several} R. Tumulka:
	Detection Time Distribution for Several Quantum Particles.
	Preprint (2016)
	\url{http://arxiv.org/abs/1601.03871}

\bibitem{detect-dirac} R. Tumulka:
	Detection Time Distribution for the Dirac Equation.
	Preprint (2016)
	\url{http://arxiv.org/abs/1601.04571}

\bibitem{Wer87} R. Werner:
	Arrival time observables in quantum mechanics.
	\textit{Annales de l'Institute Henri Poincar\'e, section A} \textbf{47}: 429--449 (1987)

\end{thebibliography}
\end{document}